\documentclass[sigconf,authorversion,pbalance=true]{acmart}

\AtBeginDocument{%
  }

\copyrightyear{2025} 
\acmYear{2025} 
\setcopyright{acmlicensed}\acmConference[ETRA '25]{2025 Symposium on Eye Tracking Research and Applications}{May 26--29, 2025}{Tokyo, Japan}
\acmBooktitle{2025 Symposium on Eye Tracking Research and Applications (ETRA '25), May 26--29, 2025, Tokyo, Japan}
\acmDOI{10.1145/3715669.3725871}
\acmISBN{979-8-4007-1487-0/2025/05}

\usepackage[capitalise,nameinlink,noabbrev]{cleveref}
\usepackage{subcaption}
\usepackage{ccicons}

\acmSubmissionID{1014}

\begin{document}

\title{Group Gaze-Sharing with Projection Displays}

\author{%
\href{https://orcid.org/0000-0003-0469-8971}{Maurice Koch},
\href{https://orcid.org/0000-0002-3310-9163}{Tobias Rau},
\href{https://orcid.org/0000-0002-4738-6655}{Vladimir Mikheev},
\href{https://orcid.org/0000-0002-5785-6788}{Seyda Öney},
\href{https://orcid.org/0000-0002-0072-1655}{Michael Becher},
\href{https://orcid.org/0009-0005-0067-2688}{Xiangyu Wang},
\href{https://orcid.org/0000-0002-6848-8554}{Nelusa Pathmanathan},
\href{https://orcid.org/0000-0003-4586-8279}{Patrick Gralka},
\href{https://orcid.org/0000-0003-1174-1026}{Daniel Weiskopf}, and
\href{https://orcid.org/0000-0003-4919-4582}{Kuno Kurzhals}} 
\affiliation{%
 \institution{\vspace{0.5em}University of Stuttgart, Germany}
 \country{}
}
\affiliation{%
  \institution{\{firstname\}.\{lastname\}@vis.uni-stuttgart.de}
  \country{\vspace{-0.5em}}
}

\renewcommand{\shortauthors}{Koch et al.}

\begin{abstract}
 The eyes play an important role in human collaboration. Mutual and shared gaze help communicate visual attention to each other or to a specific object of interest.
 Shared gaze was typically investigated for pair collaborations in remote settings and with people in virtual and augmented reality. With our work, we expand this line of research by a new technique to communicate gaze between groups in tabletop workshop scenarios.
 To achieve this communication, we use an approach based on projection mapping to unify gaze data from multiple participants into a common visualization space on a tabletop.
 We showcase our approach with a collaborative puzzle-solving task that displays shared visual attention on individual pieces and provides hints to solve the problem at hand.
\end{abstract}



\begin{teaserfigure}
\centering
  \includegraphics[width=0.57\textwidth]{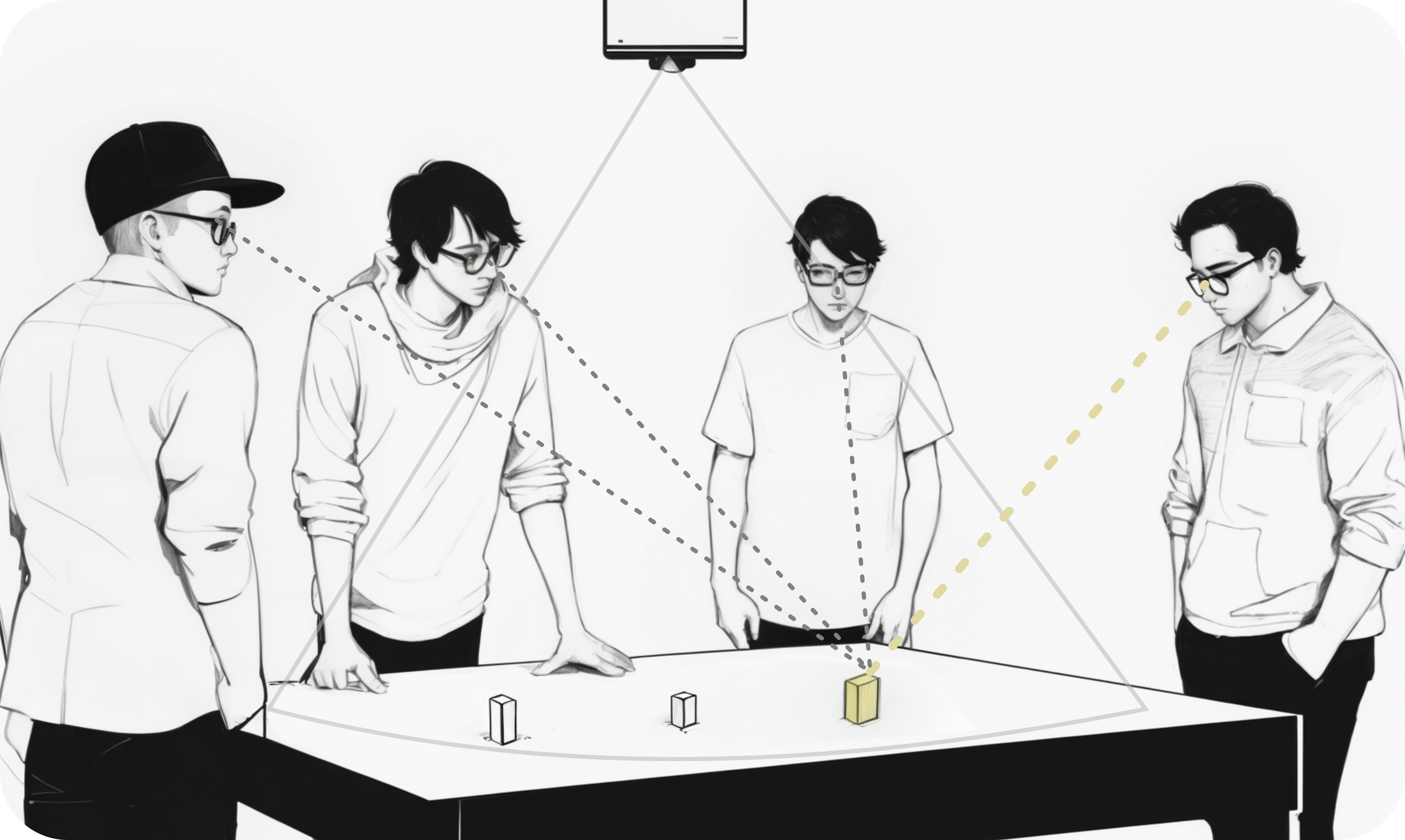}
  \caption{Projection view with four eye-tracking glasses. The gaze of each participant is tracked and mapped to the projection view on the table. This data can be used, for example, during live interaction to communicate shared attention on areas and objects of interest. (Image edited with Stable Diffusion)}
  \Description{Four people standing around a table and looking at small boxes. A projector is placed above them, indicating that the projection highlights specific objects the people look at.}
  \label{fig:teaser}
\end{teaserfigure}

\maketitle

\section{Introduction}
Social interactions in co-located scenarios such as discussions, collaborative task-solving, or just playing games with each other rely heavily on verbal and non-verbal communication.
An important part of non-verbal communication is based on mutual and shared gaze between people \cite{hessels2024gaze}.
In this work, we focus on scenarios with interactions on a tabletop because they allow for a multitude of applications with collaborations between groups of people on a single device.

Gaze cues that indicate where others are looking can be amplified in multiple ways, including visualizations in augmented reality (AR). 
While using head-mounted displays (HMDs) is currently the most common approach, projection-based AR \cite{bimber2005spatial,mine2012projection,benko2012miragetable} provides an alternative with some advantages in collaborative scenarios.
From a technical perspective, including eye-tracking glasses does not necessarily reduce the number of devices to manage.
However, it reduces the setup to a single output (the projected image), facilitating the synchronized rendering of data from all inputs (i.e., still multiple input devices in the form of the glasses).
We see the main advantage of the projection-based approach in preserving natural communication between people. 
In contrast, an HMD obfuscates eyes and faces and has to be compensated with technical tricks such as additional displays that show the face inside the HMD, as it is implemented, for instance, in the Apple Vision Pro. We suggest applying mobile eye-tracking glasses as the only wearable device to acquire gaze data. They are designed to look natural, resembling glasses people are wearing in their daily lives.
Furthermore, not all participants have to wear a device and can still see rendered content.

Our main contribution is a framework for simultaneous gaze sharing of small participant groups (tested with up to four participants) in tabletop collaboration (\cref{fig:teaser}). Our code is publicly available as a basic framework for future research on gaze-sharing applications.\footnote{This project's source code is available under: \url{https://github.com/UniStuttgart-VISUS/group_gaze_sharing}}
We showcase the technique with examples of gaze sharing on a tabletop projection.
Our approach can be used for co-located group work where the participants' points of regard are required for interactive visualization.
In the remainder of this work, we discuss further application scenarios, such as education and expert training.

\section{Related Work}

Shared gaze visualization can support group members in better understanding each other's intentions, and improve engagement.
However, designing shared gaze visualizations is challenging as the gaze is fast-moving and may exhibit erratic behavior. 
Hence, visual indicators based on raw gaze data can be highly distracting \cite{dangelo2021shared}.
\citet{atweh2025gaze} compared different real-time gaze visualizations, including gaze dots, fixation trails, and heatmaps. 
Their results indicate that the fixation trail performs best, providing a continuous visual cue of the other participants' gaze trajectories.
However, these results are context-dependent, thus not necessarily translating to other co-located settings, such as tabletop, including a group of participants.  
A comprehensive evaluation of different gaze visualizations is beyond the scope of this paper. 
Instead, we provide a first step toward investigating gaze visualization targeting co-located collaboration on tabletops.
To this end, we built upon the aforementioned studies by proposing attention grids that resemble heatmaps, and extending shared gaze visualization to physical objects.

Tabletop displays provide participants more freedom than traditional desktop computers, and can facilitate collaboration in co-located settings \cite{buisine_how_2012}. Applications include collaborative learning in classrooms \cite{prieto_studying_2014} and collaborative visual analytics \cite{isenberg_exploratory_2010}. 
Typically, they are built around displays, but they can also be realized with projection-based AR~\cite{wilson2005playanywhere, mine2012projection, bimber2005spatial, benko2012miragetable}. 
Projection-based AR can be combined with depth cameras, for example, to enable mid-air interactions using hands \cite{wilson2010combining}. 
Tabletops can be combined with free-hand \cite{benko2012miragetable} and gaze interaction \mbox{\cite{pfeuffer2016gazearchers}}.

A few works investigated gaze visualization specifically designed for tabletops.
\citet{newn2016exploring} explored gaze visualizations in tabletop games involving two players. Similar to our setup, a projector was used to augment gaze pointers in real-time in the shared gameplay area. 
In contrast, our approach is applicable to general collaborative work involving more than two participants.
\citet{pfeuffer2016gazearchers} designed a tabletop two-player game where shared and individual attention on game elements triggered specific actions. 
Our gaze trail visualization employs gamification that indicates sequential gaze behavior on tabletops of multiple participants.

\section{Application Scenarios}
We see substantial potential for group gaze-sharing in projector-based setups in many different application areas.

\paragraph{Education} In any learning environment, an instructor needs to guide attention to relevant areas on the screen~\cite{rau2024understanding}. This is typically done with a hand, finger, or pointer. Here, the gaze indicator serves as an alternative hands-free method of highlighting areas.
In addition, gaze visualization allows instructors to verify that learners are actually focusing on the intended areas.
In case a student is struggling with a task, gaze tracking can help the instructor understand and address lapses in attention, whether students are focusing on irrelevant details or missing crucial information. That could be useful in a classroom setting where an instructor needs to direct students' attention to specific areas on a screen or board, especially in subjects that require precise visual attention, e.g., mathematics, science, or art. 

\paragraph{Learning from Experts}  Multiple studies show that experts develop more efficient visual search strategies, such as in medical image inspection \cite{brunye2020eye} and chess \cite{kuchelmann2024expertise}. The nuances of visual search might be difficult to articulate in a verbal explanation, and gaze visualization could be a useful tool for this. Observing expert visual strategies can be highly beneficial for novices, and this method has already been adopted for professional training (e.g., radiology) \cite{kramer2019evaluation}. 

\paragraph{Gaze-adaptive Visualizations}
Being aware of users' visual attention behavior can help enhance their understanding of a stimulus. 
By extending the stimulus with additional visualizations or highlighting areas, attention can be guided toward under-explored regions of the stimulus~\cite{Srinivasan25}. 
Gaze adaptive visualization so far has been limited to single-user cases, and extending it to collaborative settings remains an open research area~\cite{yanez2024state}.
An example of a collaborative task that might benefit from gaze-adaptivity is geo-visualization, where participants examine climate data on a map. Gaze feedback indicates parts of the map that were already processed and guides participants to unexplored areas, for instance, by a heatmap.
Moreover, gaze-adaptive visualization can be useful for complex tasks or when users experience confusion.
Cognitive load measures or gaze transition frequencies could detect such states of confusion.

\section{Technique}
Two main steps are necessary to achieve an interactive gaze visualization for multiple participants: (1) Mapping of all sources into one common coordinate system and (2) a detection (and recognition) of objects to interact with.
Except for the required eye-tracking glasses, we propose a low-cost setup consisting of a projector and a webcam to create the tabletop environment.

\subsection{Setup}
Figure~\ref{fig:gazesetup} illustrates our setup that comprises the following hardware: consumer-grade projector, webcam, eye-tracking glasses (Pupil Labs Invisible), and a computer.
Gaze data and first-person video are streamed over Wi-Fi to the computer that performs gaze mapping onto the projection view (see \cref{subsec:mapping}).
Object detection is performed on the webcam video required for one of our three proposed gaze visualizations.
The projection view is rectangular, measuring 770\,mm in width and 550\,mm in height. 
With a projector resolution of $1900 \times 1080$ pixels, we get a pixel size of 0.405\,mm $\times$ 0.509\,mm (\Cref{fig:vis-error}). 
Technically, the setup is scalable \cite{Richer2024scalability}, considering the number of participants.
It is only limited by the available network bandwidth for streaming the gaze data and the available processing capacities of the computer. 
We evaluated the setup with four participants; current limitations are mainly based on increasing network traffic since every eye-tracking unit sends a world-view video and the respective gaze data.
A recording device that sends only mapped coordinates to the projection could significantly increase the number of participating input devices.

\subsection{Gaze Mapping} \label{subsec:mapping}
\begin{figure*}[t]
    \centering
    \begin{subfigure}[b]{0.305\textwidth}
    \centering
      \includegraphics[width=\textwidth]{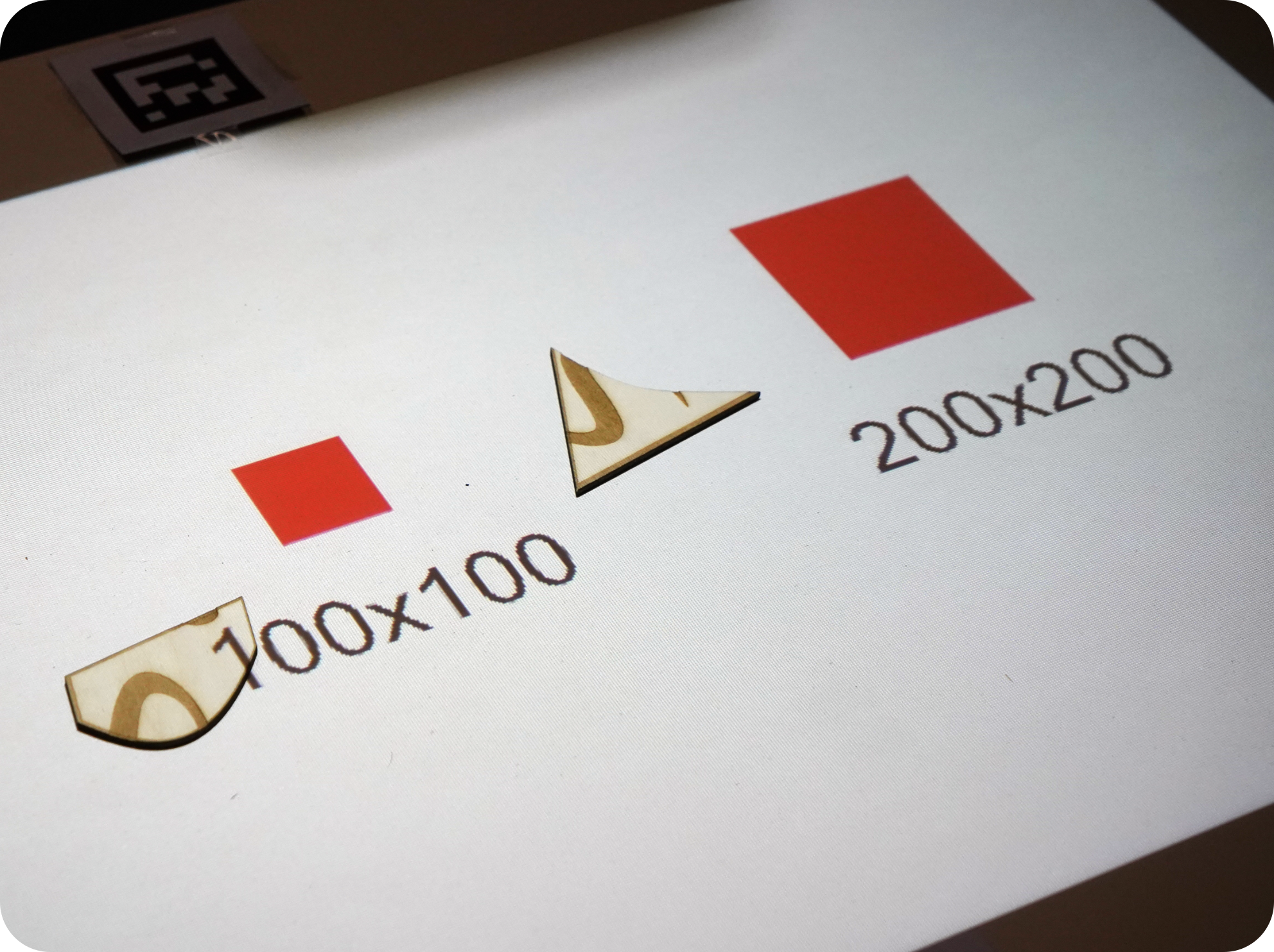}
      \caption{Relative sizes}
      \label{fig:vis-error}
    \end{subfigure}
    \begin{subfigure}[b]{0.305\textwidth}
    \centering
    \includegraphics[width=\textwidth]{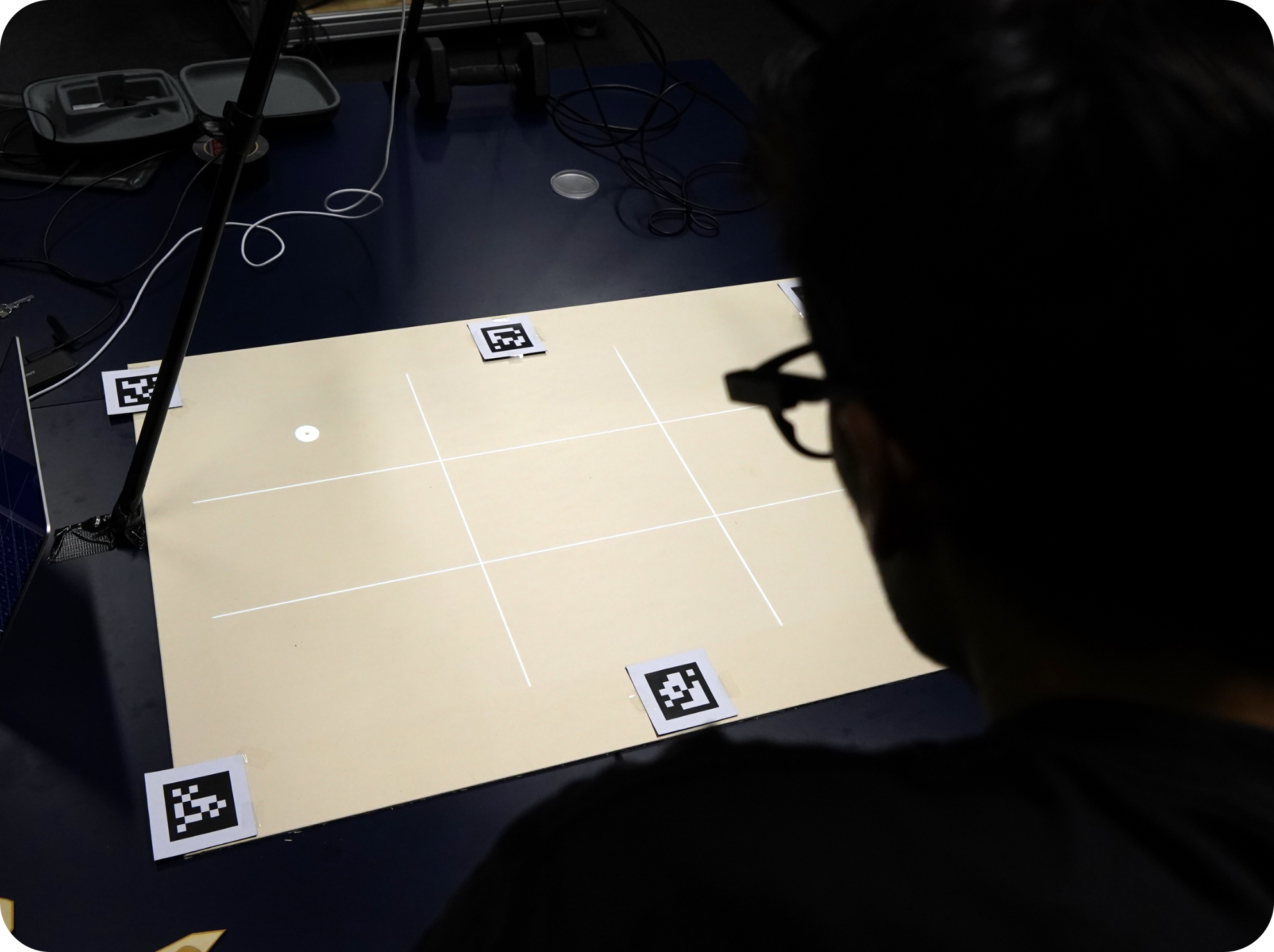}
      \caption{Horizontal view}
       \label{fig:marker-horizontal}
    \end{subfigure}
    \begin{subfigure}[b]{0.305\textwidth}
    \centering
      \includegraphics[width=\textwidth]{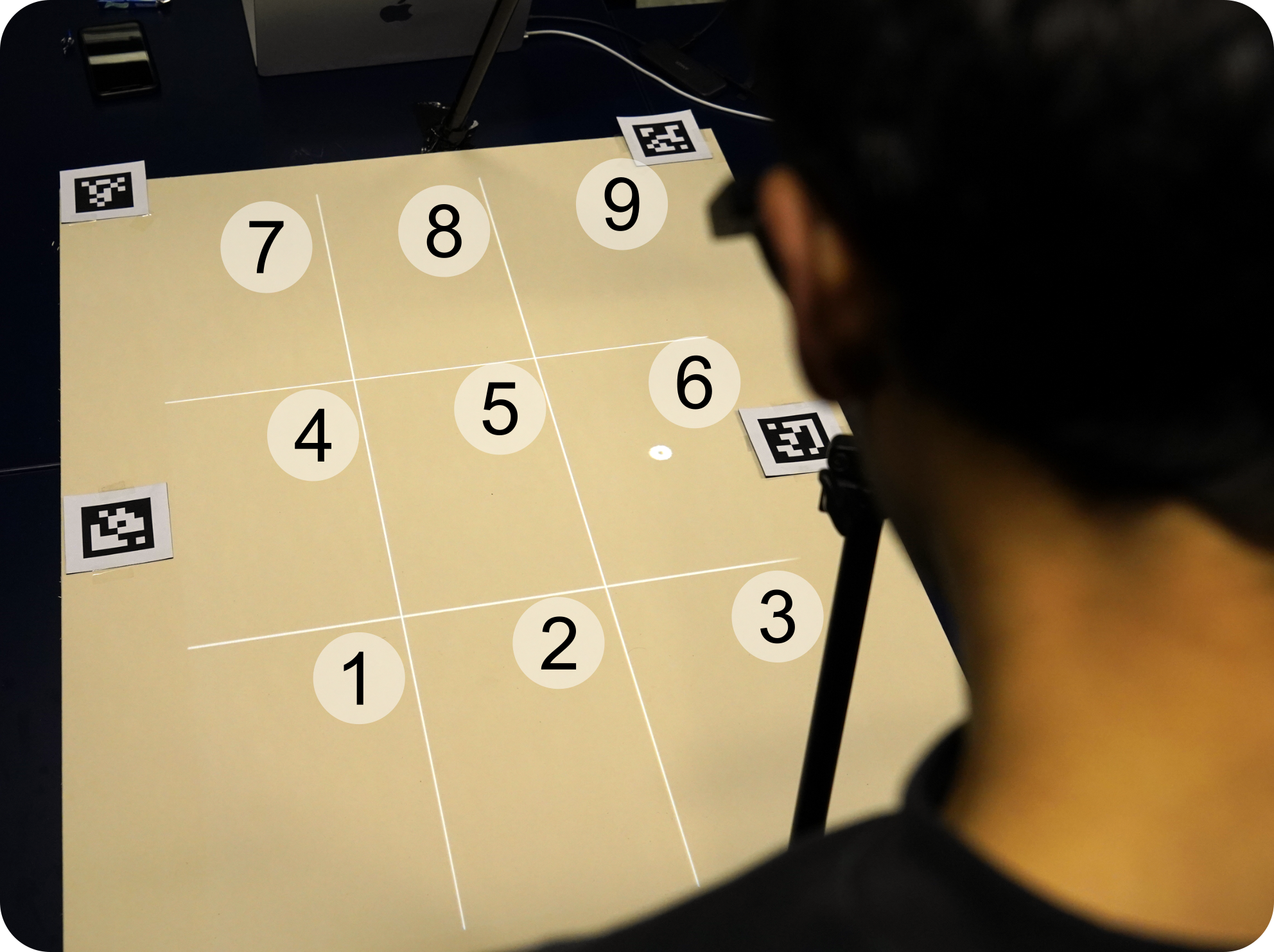}
      \caption{Vertical view}
        \label{fig:marker-vertical}
    \end{subfigure}
    \begin{subfigure}[b]{0.453\textwidth}
    \centering
      \includegraphics[width=\textwidth, trim={0 0 40 30},clip]{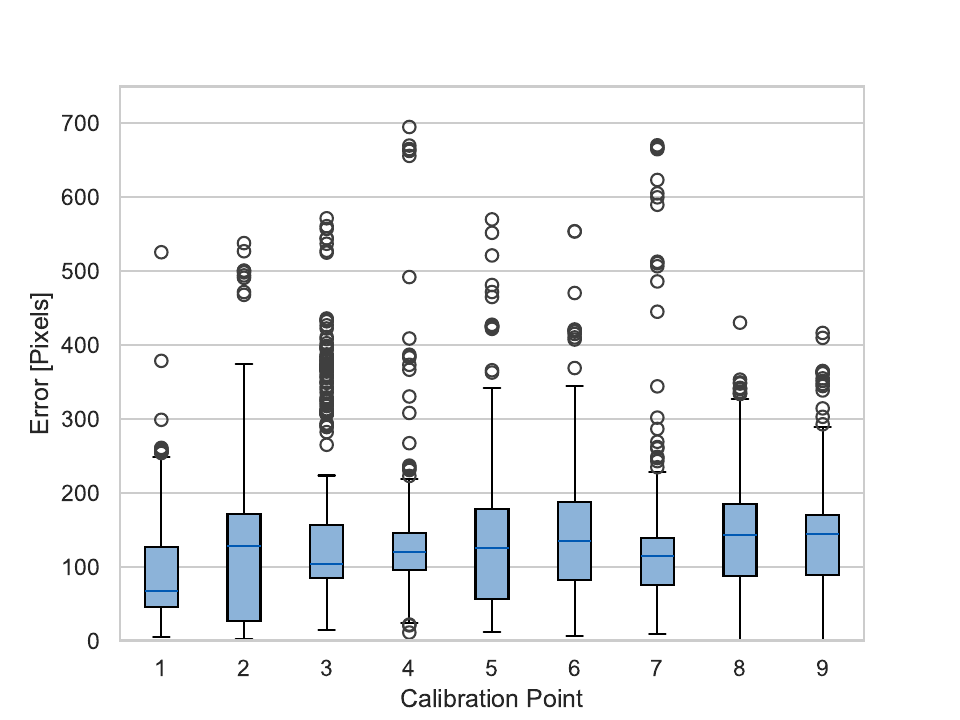}
      \caption{Error horizontal view}
      \label{fig:error-horizontal}
    \end{subfigure}
    \hspace{1em}
    \begin{subfigure}[b]{0.453\textwidth}
    \centering
      \includegraphics[width=\textwidth, trim={0 0 40 30},clip]{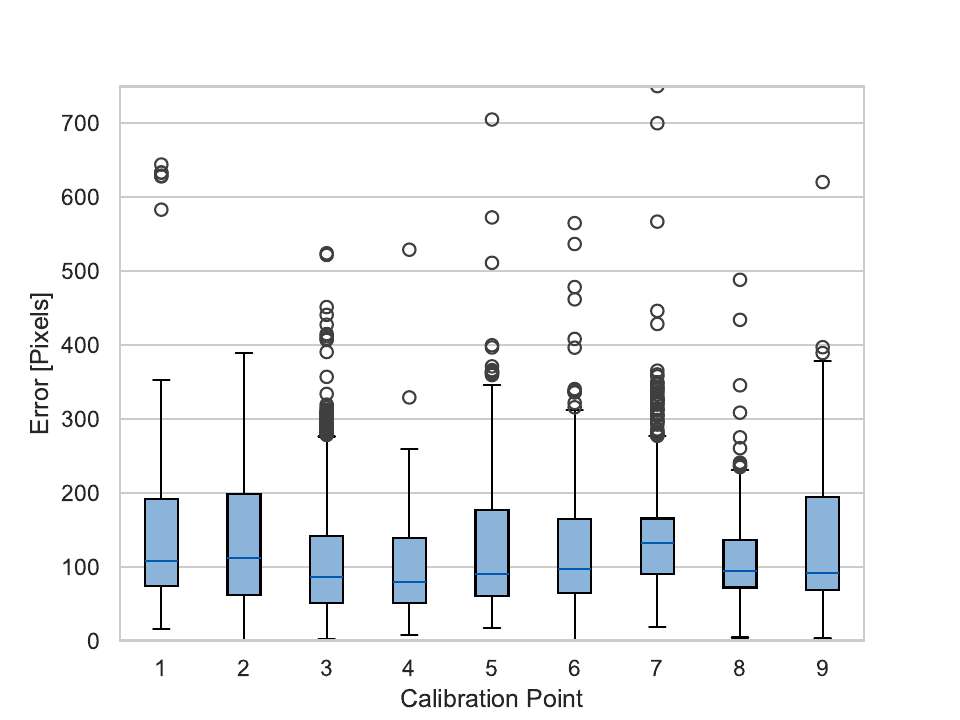}
      \caption{Error vertical view}
      \label{fig:error-vertical}
    \end{subfigure}
    \caption{Gaze mapping on the projection view. (a) shows virtual and physical objects combined to convey scale. (b) and (c) show the horizontal and vertical viewing positions used to evaluate the gaze accuracy. (d) and (e) depict the accuracy of the gaze mapping for these two viewing positions.}
    \Description{A set of images showing the accuracy measurements. Images (a)--(c) show relative sizes according to the measured accuracy and how the horizontal and vertical orientation of the projection plane was on the table.  }
    \label{fig:calibration}
\end{figure*}
The projector displays the main view to which the gaze data has to be mapped.
As shown in \cref{fig:calibration}a--c, we placed six AprilTags~\cite{Olson2011} of size 85\,mm $\times$ 85\,mm  around the projection to map the individual gaze coordinates from eye-tracking glasses to the coordinate system of the projection view.

We performed a 9-point calibration routine to evaluate the accuracy of the aforementioned gaze mapping. 
It should be noted that the purpose of our calibration routine was not to compute positional offsets.
Evaluating the robustness of gaze mapping concerning different viewing positions is particularly important for collaborative tasks that exhibit free movement.
Hence, we evaluated the accuracy from two different viewing positions: horizontal (see \cref{fig:marker-horizontal}) and vertical (see \cref{fig:marker-vertical}), both with an eye-to-screen distance of approximately 1\,m. 
Gaze samples with less than three out of six detected markers were discarded. An onset delay of 100\,ms on calibration points accounted for saccadic latency.

\Cref{fig:error-horizontal} and \cref{fig:error-vertical} show the error values for each calibration point aggregated over ten participants.
Accuracy is defined as the Euclidean distance between mapped gaze points and the calibration point.
For the horizontal view, the mean error ranges from 87 to 149 pixels across most calibration points, while the mean error for the vertical view is slightly higher, ranging from 106 to 150 pixels.
The observed differences between the two viewing positions result from varying numbers of detected markers, which impacts the fidelity of the perspective mapping.
Due to our small sample size, the aforementioned results should be considered indicative.

\subsection{Object Detection}
\label{sec:object_detection}
Simple object detection algorithms are already sufficient for most interactions, such as highlighting objects or calculating gaze intersections. 
However, specific interactions related to an individual object require recognition and tracking of each physical object in the view. 
Further, the detection must achieve at least 20 frames per second for real-time feedback.
Models like Sam2~\cite{ravi2024sam2segmentimages} provide a working detection of some objects out of the box but do not reach interactive frame rates.
A training dataset containing annotated objects is required for reliable detection. We trained our dataset with the YOLO11 model~\cite{Jocher_Ultralytics_YOLO_2023}. Detected objects are then visualized with oriented bounding boxes. 
This approach allows us to go beyond providing highlighting and simple visual hints, such as the display of neighboring pieces when users focus their attention on individual puzzle pieces for an extended time period.

\section{Examples}

\begin{figure*}[t]
    \centering
    \begin{subfigure}[b]{0.47\textwidth}
    \centering
      \includegraphics[width=\linewidth]{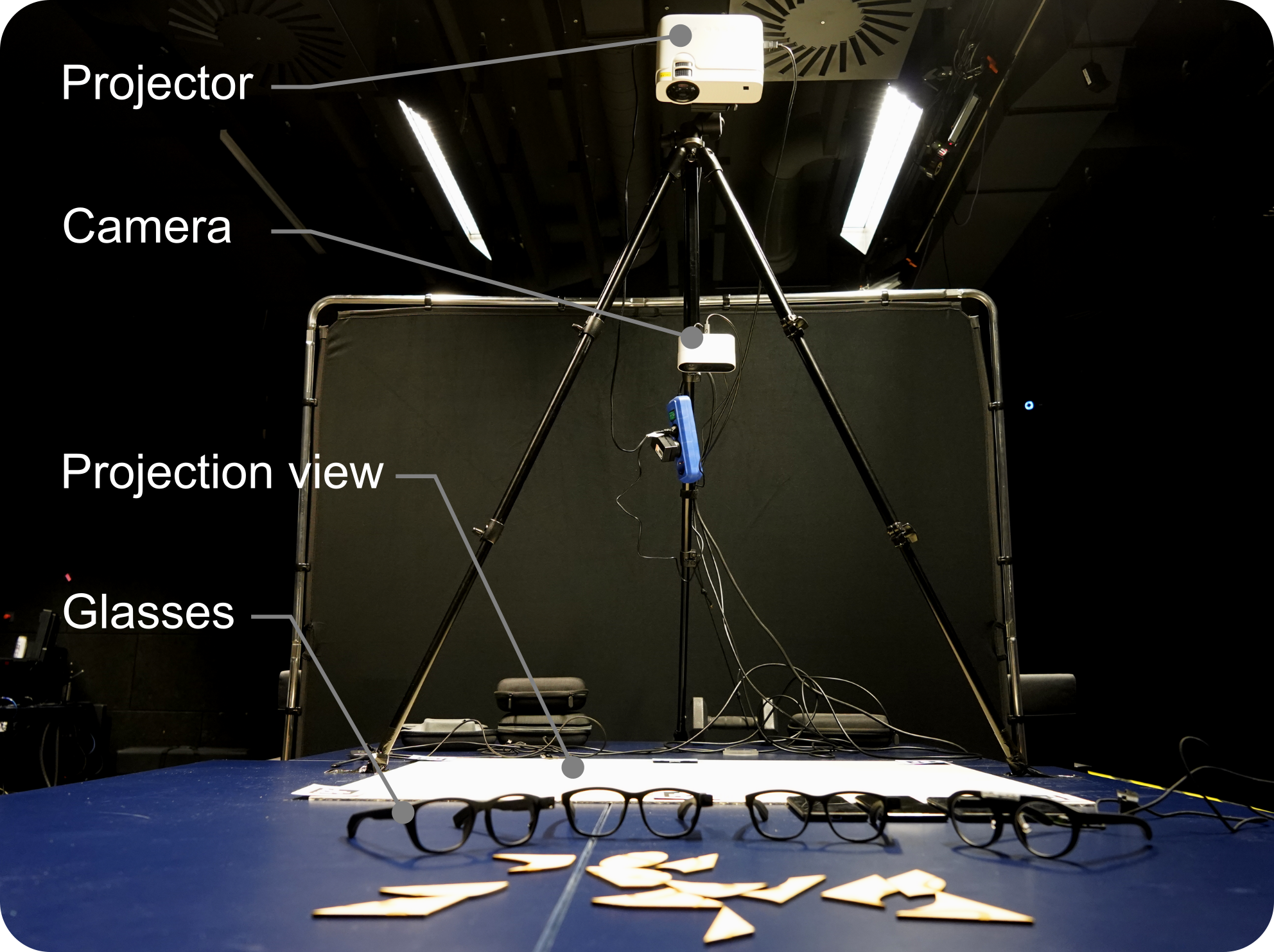}
      \caption{Hardware setup for gaze sharing}
      \label{fig:gazesetup}
    \end{subfigure}
    \hspace{4.5ex}
    \begin{subfigure}[b]{0.47\textwidth}
    \centering
      \includegraphics[width=\linewidth]{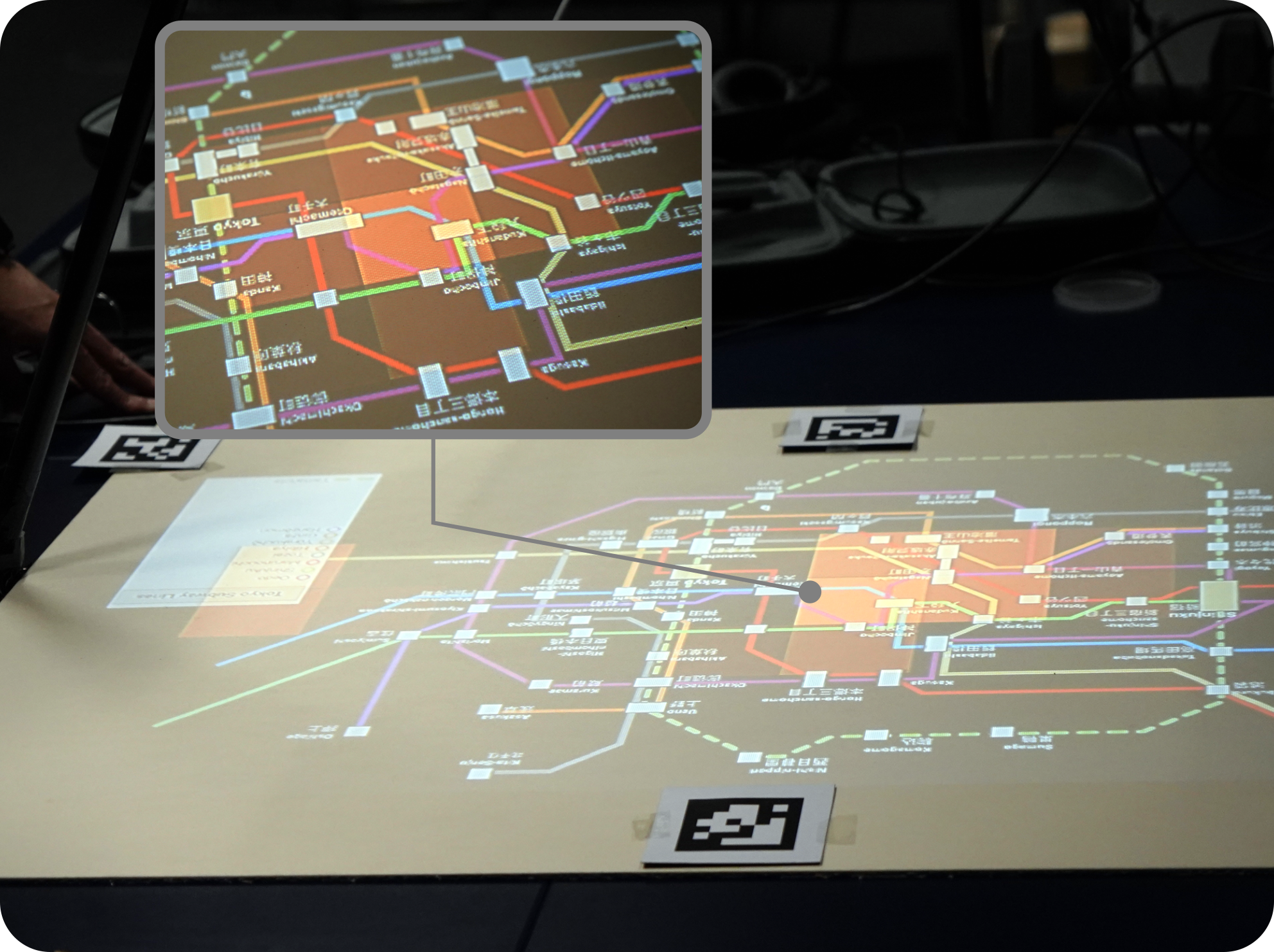}
      \caption{Shared heatmap visualization on a map}
      \label{fig:sharedgaze}
    \end{subfigure}
    \begin{subfigure}[b]{0.47\textwidth}
    \centering
      \includegraphics[width=\linewidth]{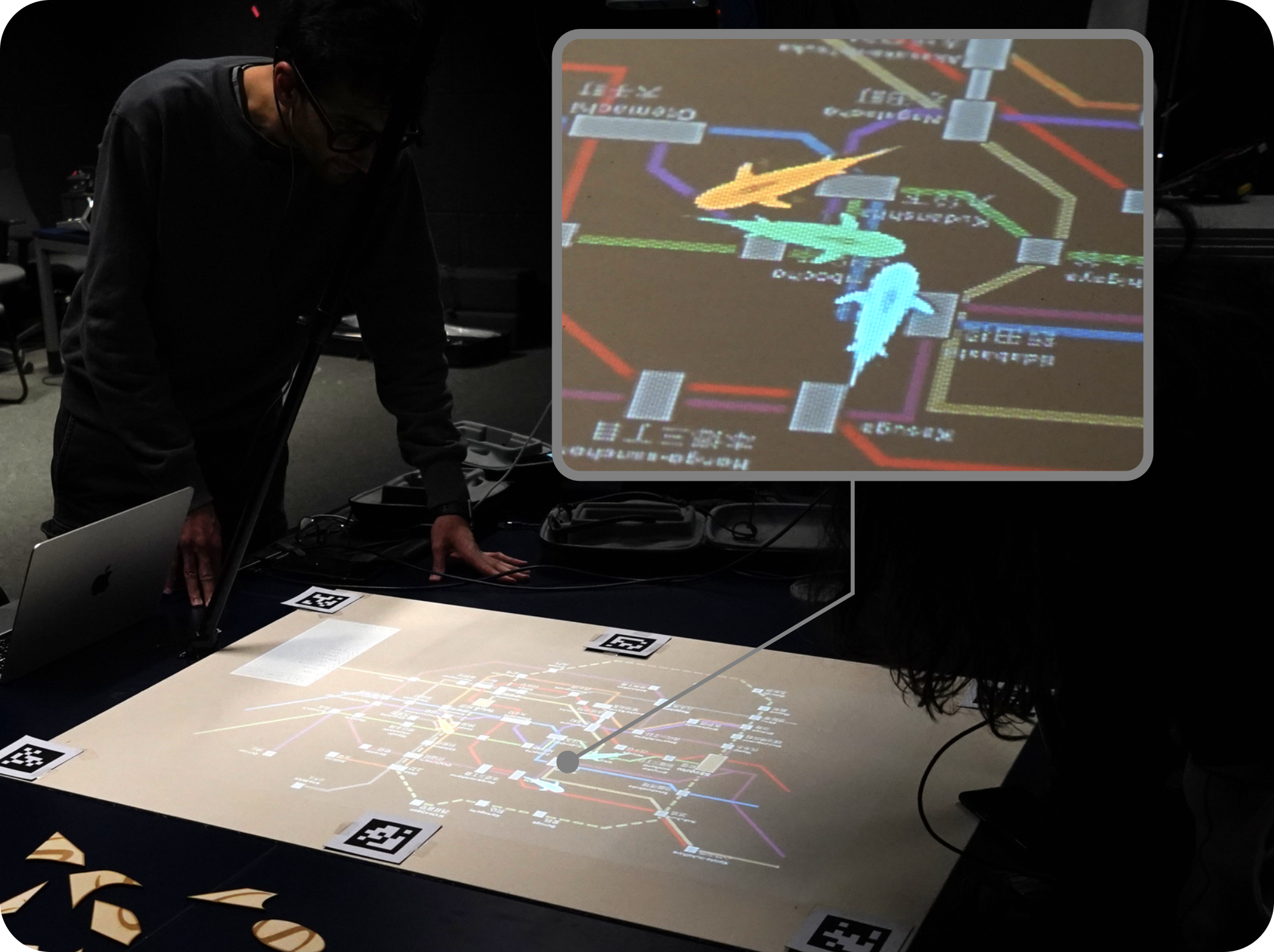}
      \caption{Gaze trails visualized by sharks following the gaze\vspace{-2ex}}
      \label{fig:gazetrails}
    \end{subfigure}
    \hspace{4.5ex}
    \begin{subfigure}[b]{0.47\textwidth}
    \centering
    \vspace{1ex}
      \includegraphics[width=\linewidth]{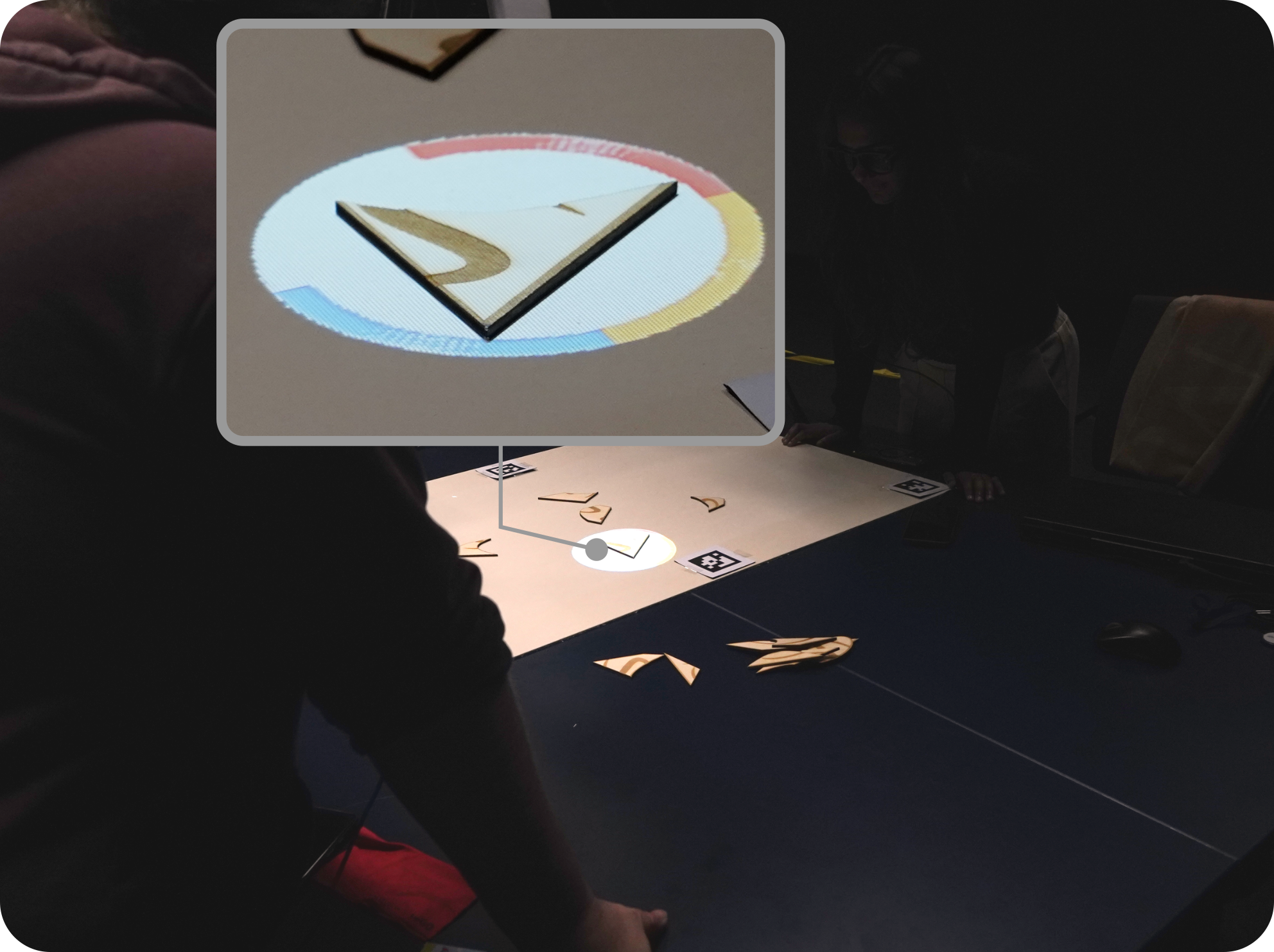}
      \caption{Example of physical objects (puzzle pieces) in the projection\vspace{-2ex}}
      \label{fig:puzzle}
    \end{subfigure}
    \caption{Implemented shared gaze examples. (a) The setup includes a camera and a projector to visualize gaze from multiple participants as (b) a grid-based heatmap, (c) gaze trails by animated virtual objects, and (d) highlights on physical objects.\vspace{0.5ex}\\ 
    \scriptsize (Map in (c--d): \textcopyright~ Comicinker CC BY-SA 3.0: \url{https://commons.wikimedia.org/wiki/File:Tokyo_subway_map_en_jp.svg}.
    Shark sprite: \textcopyright~ Aurora CC BY 4.0: \url{https://aurora-sprites.wixsite.com/main}).
    Images (a)--(d): \textcopyright~ Koch et al. CC BY-SA 4.0
    }
    \Description{Four images of the implemented setup. (a) Shows a projector and a camera on top, pointing at a table. (b) Shows a metro map projection enriched by a heatmap. (c) Shows the same map with shark sprites following the users' point of regard. (d) It shows a puzzle piece enriched by a projection circle that shows how many users currently look at the piece.}
\end{figure*}
To showcase different examples of gaze-sharing with our proposed method, we prepared three example tasks that require the collaboration of the participants. Our examples are written in Python.

\subsection{Grid-based Heatmap}
We employ AOI grids that divide the table into equal-sized cells that aggregate dwell time (\emph{attention grids}). Depending on the task requirements, the grid may be displayed.
For each grid cell, we employ an upper bound on the dwell time, which increases the saturation. 
Additionally, we implement decay over time when previously highly attended areas are no longer the focus. This dynamic adjustment causes accumulated dwell times to fade out.
Attention grids go beyond puzzle solving and apply to various collaborative scenarios. For example, it can aid in visual search tasks (e.g., ``Where’s Waldo?'') or assist an expert in demonstrating gameplay strategies to a novice in activities such as table football or chess.

Joint attention is essential to highlight which areas of the table have been thoroughly examined and which areas received less attention. To facilitate this, we visualize the values of a single attention grid that accumulates dwell times across all participants in a heatmap (see \cref{fig:sharedgaze}). 
Each grid cell is assigned a color according to its accumulated dwell time to indicate joint attention on the table. A cell remains uncolored until it is fixated upon. After exceeding a predefined dwell time threshold, the cell adopts a low-value color on a continuous single-hue color scale. With prolonged gaze, the color gradually shifts along the scale (e.g., toward red), reflecting increased joint attention.

This grid-based heatmap forms an artificial \textit{inhibition of return} (IOR)---a cognitive mechanism that discourages reorienting attention to previously attended locations. IOR makes it less likely for attention to return to the most salient areas, encouraging observers to scan other regions of the visual field. The \textit{collective IOR} can be beneficial for gaining new insights and ensuring a more comprehensive analysis of the environment.

\subsection{Gaze Trails}
As mentioned by \citet{atweh2025gaze}, a trail visualization provides visual cues regarding a participant's gaze trajectory, which fosters insights into their cognitive processes within a co-located setting. We assign each participant an individual virtual grid and trail that follows the participant's gaze toward the cell position within the grid with the highest dwell time.  
To ensure that the collaborators are following the gaze trail of a participant, we visualized a shark \textit{swimming} along the trajectory of the participant's gaze to display the trail (see \cref{fig:gazetrails}).
This provides a potential for gamification, making the attention guidance more engaging. We could add trailing waves following the shark's movement to highlight the trajectory. 
In addition to displaying the trails, our visualization also highlights areas of joint attention, similar to a grid-based heatmap. However, rather than intensifying a cell’s color, joint attention is indicated when multiple sharks occupy the same grid cell.

\subsection{Physical Object Highlighting} 
In data physicalization, tangible objects directly represent data, transforming abstract information into a physical form~\cite{Jansen15}. 
This makes the data tangible and enables intuitive interaction, facilitating deeper understanding and exploration of the data while intensifying user engagement.
However, integrating multiple tangible objects can make their identification and manipulation more challenging. 
This challenge becomes evident in a puzzle-solving task that involves placing, connecting, and orienting several objects on the tabletop. 
This task requires a lot of experimenting (e.g., taking two pieces and seeing if they fit), while in a real-world scenario, multiple viable outcomes can be expected. 
Gaze cues on the physical objects provide a rough insight into the thinking process of collaborators.
Similar to \citet{rau2024understanding}, we use a circular highlight and arc segments around it to display individual attention momentarily on an object (see \cref{fig:puzzle}).

As described in \cref{sec:object_detection}, we fine-tuned the YOLO11 model to perform real-time detection of our puzzle pieces.
To this end, we created a dataset comprising 56 manually annotated pictures of our puzzle pieces.
To augment our dataset, we used common image transformations, flip, crop, shear, grayscale, brightness, and blur. In total, we collected 168 pictures for fine-tuning.
The training result yields an accuracy of approximately $70\%$, which we then temporally smooth by buffering the last 20 detections and picking the one with maximum confidence.
Generally speaking, we trade longer reaction time when moving puzzle pieces for better detection stability.

\section{Discussion}
In this section, we discuss technical considerations and potential directions for future research.

\paragraph{Out-of-Frame Detection}
To this point, we perform all object-related detections with the external webcam.
This setup restricts the detection to objects in the projection view. We plan to include an analysis of the individual video streams from the eye-tracking glasses. This extension would allow us to support additional tasks, for example, when participants moved objects out of the projection view to work on them individually. Furthermore, detecting faces in combination with gaze would allow quantifying and visualizing mutual gaze between participants, opening numerous opportunities to communicate visual attention during collaboration.

\paragraph{Mixed Audience Participation}
We see the proposed setup's main strength in its flexibility for participants.
To emphasize this point, not all participants must wear eye-tracking glasses to contribute to collaborative tasks. The support of gaze-based functions aims to better communicate visual attention and help highlight objects in the focus of discussion.
Hence, it would also be possible to dynamically switch who is wearing the glasses, depending on their need to communicate where they are looking. 

\paragraph{Post-experiment Analysis}
Our current implementation focuses on live interaction and visualization.
The post-experimental analysis of recorded scan paths \cite{Privitera2000} provides potential new insights into collaborative behavior in general. Such an analysis would require the recording of data and the application of established eye-tracking metrics and visualization techniques \cite{Blascheck17}. While the recording is not much more effort to include, new techniques will be required to extract the necessary information from the data depending on the research question, including new visualization methods for visualization research (Vis4Vis) \cite{Weiskopf2020Vis4Vis}.

\paragraph{Future Studies}
Our preliminary results regarding the accuracy of the implemented approach are promising and require further validation by future experiments. We plan to investigate different angles, distances, and an application to long-term education and training scenarios. Further, the upper boundary of supported participants has to be tested in the future.

\section{Conclusion}
We presented a framework for developing new interaction and visualization techniques for gaze in group collaborations.
We see much potential in using shared gaze to support communication-intensive scenarios such as teaching, training, and collaborative design. 
There are opportunities to extend this technique to design gaze-responsive systems that react not just to individual participants but also consider the gaze distribution of the entire group.
The use of projection-based displays allows for unconstrained face-to-face interaction between people and is more lightweight than HMDs for all participants.
While interaction with the virtual content requires wearing eye-tracking glasses, participation in the tasks is more inclusive and allows also people without additional hardware to join discussions at the tabletop.
We see the presented approach as a framework for developing new ways to explore interactive visualizations for group collaboration. 
The use of projection displays is especially fit for workshop scenarios that foster group collaboration in combination with physical objects.

\section*{Ethics and Privacy}
Our paper is concerned with general issues of ethics and privacy of eye-tracking research. In particular, communicating your gaze to others in our collaborative scenarios reveals personal information. We focus only on gaze on the workspace, involuntary recording of gaze outside the projection area is not processed.

\begin{acks}
This project started at the SFB-TRR 161 Hackathon 2025. This work was funded by \grantsponsor{DFG}{Deutsche Forschungsgemeinschaft (DFG, German Research Foundation)}{https://www.dfg.de/} under Germany’s Excellence Strategy - EXC 2075 -- \grantnum{DFG}{390740016} (SimTech), EXC 2120/1 -- \grantnum{DFG}{390831618} (IntCDC), Project-ID \grantnum{DFG}{449742818}, \grantnum{DFG}{Project-ID 279064222} -- SFB 1244, and Project-ID \grantnum{DFG}{251654672} -- TRR 161; and by the \grantsponsor{BMBF}{German Federal Ministry of Education and Research (BMBF)}{https://www.bmbf.de/EN/Home/home_node.html} through the Alexander von Humboldt Foundation and Project-ID \grantnum{BMBF}{16ME0608K} (WindHPC).
\end{acks}

\bibliographystyle{ACM-Reference-Format}

\end{document}